# Attosecond Path Qubits in Strong-Field Physics

Oren Cohen

Solid State Institute, Physics Department and Helen Diller Quantum Center, Technion-Israel Institute of Technology, Haifa 3200003, Israel

*Physics Department, Guangdong Technion – Israel Institute of Technology, Guangdong 515063, China*

## Abstract

This perspective introduces attosecond path qubits: measurement-defined two-level subsystems that arise naturally in strong-field physics from the coherent interference of distinguishable quantum pathways. These effective qubits are dynamical superposition states formed during laser-driven sub-cycle electron motion and appear in diverse settings, including short-long trajectories in high-harmonic generation (HHG), early-late ionization bursts in multicolor fields, polarization-selective recombination channels, and momentum or energy-bin interference in above-threshold ionization (ATI). Attosecond qubits can be prepared, coherently manipulated, and read out using existing attosecond techniques, providing a compact Hilbert-space description of strong-field dynamics beyond semiclassical trajectory pictures. A density-matrix formulation makes explicit the roles of coherence, dephasing, and decoherence, arising from continuum propagation, Coulomb interaction, infrared dressing, and macroscopic averaging, within a reduced description that remains far from thermalization. The attosecond qubits framework identifies new opportunities for ultrafast coherence metrology and for probing open quantum dynamics in strongly driven, time-dependent Hamiltonians.

## 1. Introduction

Strong-field and attosecond physics provide direct access to the natural timescale of electronic motion, where tunneling, sub-cycle acceleration, and recollision unfold on tens to hundreds of attoseconds [1-9]. These processes create coherent superpositions of electronic pathways that shape observables ranging from HHG spectra to momentum-resolved ATI distributions. Although the underlying physics of the electronic motion is inherently quantum, many descriptions rely on semiclassical approximations that interpret these pathways as classical trajectories. This view explains a wide range of observables, but it does not provide an explicit identification of the quantum subsystems, basis structures, and coherence or decoherence channels associated with a given measurement in strong-field dynamics.

Here we propose that pairs of strong-field pathways contributing coherently to a measurable observable can be identified as path qubits [10,11], and term them attosecond path qubits. The notion of an attosecond path qubit provides a compact Hilbert-space representation of strong-field interference and offers a unified framework for describing coherence, control, and decoherence in strong-field dynamics. The approach is operational: pathway states become distinguishable and acquire meaning as basis states through standard strong-field measurements such as harmonic spectra, angular emission patterns, or momentum-resolved electron detection. This framing allows strong-field interference to be discussed using qubit language: coherence, phase control, and decoherence.

Previous work in high-harmonic generation (HHG) has demonstrated that strong-field observables can be decomposed into contributions from distinct quantum trajectories, such as short and long recollision paths [1,2], and that these contributions can be selectively accessed through far-field divergence [12], polarization [13], and phase-sensitive interferometric measurements [4–7]. In parallel, theoretical quantum-optical descriptions of HHG have reframed strong-field emission in terms of quantum electrodynamics and quantum states of light and matter [14,15], while experimental trajectory selectivity established HHG as a sensitive probe of sub-cycle electronic coherence. Also, density-matrix formalisms have been employed in attosecond and strong-field physics to address specific reduction mechanisms [16-23]. However, previous works, whether focused on trajectory-resolved observables, on the quantum state of the emitted field, or on particular microscopic subsystems have not identified the effective electronic quantum subsystems defined by the measurement itself as path qubits. The attosecond-qubit framework introduced here builds on these advances by providing a measurement-defined Hilbert-space description that unifies trajectory interference, coherent control, and decoherence within a compact quantum framework.

## 2. Attosecond Path Qubits: Definition and Examples

Strong-field dynamics routinely generate coherent superpositions of quantum pathways that contribute to a given observable. An attosecond qubit is defined when such dynamics can be operationally restricted to an effective two-dimensional Hilbert subspace associated with a specific measurement. Importantly, this qubit is a measurement-defined quantum object: the qubit basis acquires meaning only with respect to an observable that distinguishes and interferes contributions from two dominant pathways.

Consider a strong-field electronic state evolving under a time-dependent laser-driven Hamiltonian. For a chosen observable, such as a given harmonic order, emission direction, polarization channel, or photoelectron momentum bin, the detected signal often receives dominant contributions from two quantum pathways. The electronic state projected onto the subspace relevant for this measurement can then be expressed as

$$|\psi\rangle = c_1 e^{i\frac{S_1}{\hbar}} e^{i\sigma_1}|1\rangle + c_2 e^{i\frac{S_2}{\hbar}} e^{i\sigma_2}|2\rangle \quad (1)$$

where $|1\rangle$ and $|2\rangle$ label pathway-associated contributions selected by the measurement, $S_{1,2}$ are the corresponding strong-field actions, $c_{1,2}$ encode coherent amplitudes determined by pathway-dependent coupling matrix elements relevant to the chosen observable, and $\sigma_{1,2}=\alpha_{1,2}+i\beta_{1,2}$ represent stochastic complex phase contributions. In the standard strong-field approximation, the saddle-point action is complex even for a single atom driven by a perfectly defined laser field; at this level the electronic dynamics remain fully coherent, with the complex action determining relative phases and amplitudes of different quantum pathways within an overall pure quantum state. Dephasing and decoherence emerge upon projection onto a reduced, measurement-defined subspace [footnote 24].

Although $|1\rangle$ and $|2\rangle$ are generally neither orthogonal nor stationary in the full electronic Hilbert space, they form an effective two-level system with respect to the chosen observable. Operational orthogonality is established when the measurement distinguishes the two contributions sufficiently to yield high-visibility interference. In this effective two-level subspace, the coefficients of the density matrix obtained by tracing over electronic and photonic degrees of freedom not resolved by the measurement and subsequently averaging over shot-to-shot and ensemble fluctuations are given by: $\rho_{jj}=|c_j|^2\langle\exp(-2\beta_j)\rangle$ and $\rho_{jk}=c_j c_k^* \exp(i(S_j-S_k)/\hbar)\Gamma_{jk}$ where $\Gamma_{jk}\equiv\langle\exp[i(\sigma_j-\sigma_k)]\rangle$ and $\langle\cdot\rangle$ denotes averaging over the fluctuations. Physically, a decreasing $\Gamma_{jk}$ reflects reduced coherence in the effective two-path subsystem due to both averaging over classical shot-to-shot fluctuations and entanglement with degrees of freedom that are not measured (e.g. continuum electron motion, residual ionic states [17,18], and emitted radiation), even though the electron dynamics is attosecond-scale, hence it does not experience thermalization [25].

A prototypical attosecond qubit consists of the short and long quantum trajectories in HHG [2,12]. For a given harmonic order, stationary-phase analysis identifies two dominant recollision pathways associated with distinct ionization and recombination times. Their action difference determines the relative phase, while recombination dipoles set the amplitudes. Measured harmonic intensities, phases, divergences, and polarization states thus correspond to projections of an underlying electronic attosecond qubit, defined with respect to the selected harmonic and detection geometry.

Time-domain qubits arise in multi-color driving fields. In cross-polarized ω-2ω configurations, two well-separated ionization bursts occur within a single optical cycle [26]. These early and late emission windows define a temporal, time-bin electronic qubit, with coherence encoded in the difference of their strong-field actions. Adjusting the relative phase between the driving fields rotates the qubit by modifying the pathway phase difference, while polarization-resolved HHG detection provides access to the corresponding Bloch-vector components.

In ATI [9], momentum- or energy-bin structures similarly reflect interference between two dominant saddle-point ionization pathways. Selecting two angular or radial momentum bins isolates a coherent two-level subsystem whose phase evolution is determined by continuum propagation and Coulomb-induced phase shifts. Such ATI qubits provide a direct probe of ionization-time coherence without requiring recombination, offering complementary sensitivity to continuum-induced phase dispersion and strong-field decoherence mechanisms.

While attosecond qubits are fundamentally electronic in origin, HHG provides a coherent interface between electronic pathway superpositions and the emitted radiation field. At the microscopic level, HHG maps electronic coherence onto photonic degrees of freedom on sub-cycle timescales. The emission process therefore correlates electronic pathways with specific photonic modes, and the joint electron-photon state generally exhibits light-matter correlations associated with the underlying pathway structure [14,15].

In standard HHG experiments driven by classical fields and simple media, these microscopic correlations are strongly diluted in conventional observables [14,15]. Coherent emission from many emitters dominates the signal, and the radiation field is well described by classical amplitudes reflecting mean electronic coherence. As a result, correlations inherited from electronic pathway interference are typically hidden in intensity-based measurements. These correlations may become accessible under suitable conditions, e.g. nonclassical driving fields or fluctuation-based, conditional, or multi-mode detection schemes. Thus, photonic pathway qubits may arise as effective two-level subsystems defined by coherent superpositions of radiation modes associated with distinct electronic pathways. Here, the notion of a photonic qubit refers to a pathway-defined mapping of electronic coherence and noise onto radiation modes, rather than to a field-level quantum-optical description of HHG.

## 3. Operations and Decoherence in Attosecond Path Qubits

Qubit operations in strong-field systems emerge naturally from standard waveform-control techniques [1-9]. Adjusting the carrier-envelope phase modifies ionization phases and strong-field action differences, resulting in deterministic phase shifts of the off-diagonal coherence between pathways and thus implementing rotations of the attosecond qubit on its Bloch sphere. Changing the ellipticity or polarization of the driving field modifies the relative coupling strengths of different pathways, effectively rotating qubits defined by polarization- or symmetry-selective channels. Tailored multicolor fields can impose controlled phase differences between pathways associated with distinct ionization or emission windows, enabling controlled phase shifts analogous to single-qubit phase gates. Weak tagging or perturbative fields selectively affect one pathway and provide additional control knobs for qubit rotations within the measurement-defined subspace.

Within the density-matrix description introduced above, these operations primarily act on the off-diagonal coherence element $\rho_{jk}$. Deterministic strong-field dynamics modify its phase through changes in the action difference $S_j-S_k$, while the populations $\rho_{jj}$ are controlled by pathway-dependent coupling amplitudes. In this sense, many strong-field control protocols correspond to coherent, effectively unitary transformations of the effective two-level system, even though the underlying dynamics involve highly nonstationary continuum states.

Decoherence in strong-field dynamics arises from various physical mechanisms. Wavepacket spreading in the continuum reduces the spatial and temporal overlap between different pathways, leading to momentum- and energy-dependent phase dispersion. Similarly, Coulomb interaction induces pathway-dependent phase shifts that vary across the detected ensemble. These effects primarily cause dephasing: they modify the relative phase of the off-diagonal coherence element $\rho_{jk}$ in a pathway- and momentum-dependent manner and reduce interference visibility under ensemble averaging, while the global electronic dynamics remain coherent and unitary.

Genuine decoherence of attosecond qubits, on the other hand, arises because the electronic pathways get entangled with degrees of freedom that are not measured [25]. For example, the emission of low-energy infrared photons during strong-field evolution produces pathway-dependent radiation dressing. Tracing over these photonic degrees of freedom reduces the magnitude of the coherence factor $\Gamma_{jk}$. Similar entanglement-induced decoherence can arise from correlations with continuum electron motion outside the selected two-path subspace, residual ionic states, or multielectron dynamics.

Together, these mechanisms determine the purity and coherence of attosecond qubits in strong-field systems. The coexistence of controllable unitary phase evolution, reversible dephasing due to ensemble averaging, and entanglement-driven decoherence in a reduced description highlights attosecond qubits as intrinsically open yet non-thermal quantum systems. The electronic dynamics unfold on sub-cycle timescales that are far shorter than any thermalization or equilibration processes, making attosecond qubits a natural platform for studying ultrafast open quantum dynamics in strongly driven, time-dependent Hamiltonians without invoking dissipation.

## 4. Multi-Qubit Structures

Strong-field processes often give rise to more than one effective qubit, reflecting the coexistence of multiple distinguishable pathway pairs within a single experiment. In most cases, these structures correspond to (possibly) correlated projections onto different observables.

In HHG, each harmonic order represents a distinct projection of the same microscopic electronic attosecond qubit, defined by interference between short and long trajectories. The resulting set of harmonic modes forms a ladder of related two-level systems distributed across the harmonic plateau. Correlations between adjacent harmonic orders provide information on how microscopic coherence encoded in the off-diagonal density-matrix elements propagates across recollision energies and emission frequencies.

Additional effective qubits arise from spatial and modal degrees of freedom. In non-collinear or multi-beam geometries, different angular emission branches correspond to distinct projections of the electronic qubit, while the spatial mode structure of the emitted radiation defines an additional macroscopic degree of freedom. In low-photon-number or mode-resolved regimes, spatially separated HHG modes may form a photonic path qubit whose amplitudes reflect the underlying electronic coherence. Such configurations give rise to hybrid microscopic–macroscopic pathway structures where electronic and photonic pathway qubits coexist but are not generically independent.

Higher-dimensional structures also emerge naturally when more than two pathways contribute coherently to a given observable. Bicircular $\omega$-$2\omega$ driving fields [27] generate three well-separated attosecond emission bursts per optical cycle, corresponding to an electronic qutrit defined by three interfering ionization-recollision pathways. More complex waveform synthesis can produce larger sets of distinguishable emission windows, enabling controllable electronic qudits whose dimensionality is determined by the measurement resolution rather than by stationary level structure.

Together, these multi-qubit and higher-dimensional configurations illustrate that strong-field systems naturally support rich, measurement-defined Hilbert-space structures. While true tensor-product architectures are not generic, correlated qubit and qudit projections across electronic, spatial, and photonic degrees of freedom provide access to coherence and correlations spanning multiple scales, from sub-cycle electronic motion to macroscopic radiation modes.

## 5. Outlook

The attosecond-qubit framework introduced here provides a compact Hilbert-space description of strong-field electron dynamics, identifying measurement-defined two-level subsystems embedded within sub-cycle interference phenomena. By formulating strong-field pathways as effective qubits, this approach establishes a common language for coherence, control, and decoherence across HHG, ATI, and related processes.

The density-matrix description highlights attosecond strong-field systems as settings for studying ultrafast open quantum dynamics. Coherent waveform control enables manipulation of qubit phases and populations, while continuum propagation, Coulomb interaction, infrared dressing, and macroscopic averaging introduce dephasing and decoherence on attosecond timescales. In this context, density-matrix-based spectroscopy provides indirect experimental access of entanglement between selected electronic pathways and unmeasured degrees of freedom, through its imprint on coherence, purity, and correlated fluctuations in regimes far from equilibrium and under strongly time-dependent Hamiltonians. Viewed through a quantum-thermodynamic lens [28], this framework further enables experimental access to information-theoretic entropy growth associated with entanglement and coarse-graining in a reduced description, rather than with heat flow or equilibration, thereby probing pre-thermal irreversibility during coherent, strongly driven electron dynamics on sub-cycle timescales.

Beyond single effective qubits, strong-field processes support correlated multi-qubit and higher-dimensional structures defined by electronic, spatial, and photonic pathway degrees of freedom. While such structures are generally not tensor-factorized subsystems, their correlated projections encode coherence and correlations spanning microscopic electronic motion and macroscopic radiation modes. In this context, the photonic pathway-qubit structure identified here suggests routes toward extreme ultraviolet radiation with non-classical properties encoded in correlations and fluctuations. Such states may enable enhanced attosecond spectroscopy and metrology schemes that exploit quantum correlations while operating in regimes of high photon flux and strong-field dynamics. In this sense, strong-field systems offer a form of analog quantum simulation in which tunneling prepares coherent superpositions, continuum dynamics implements Hamiltonian evolution, and detection acts as a projective measurement on attosecond qubits.

More broadly, attosecond qubits position strong-field physics within the landscape of quantum science, not as a scalable quantum-information platform, but as a uniquely fast and controllable setting for exploring coherence, dephasing, and decoherence at the fundamental timescale of electronic motion. This perspective may inspire new directions for probing quantum dynamics in extreme, ultrafast, and strongly driven regimes that are inaccessible in conventional quantum systems.

Decoherence arises only when fluctuations of the action, whether in its real or imaginary components, are correlated with electronic, ionic, or radiative degrees of freedom that are not measured and are traced out in the reduced description. In such cases, pathway-dependent attenuation encoded in the imaginary action often parametrizes the loss of coherence by reflecting which-path information carried by unobserved degrees of freedom, but it is the act of tracing over these degrees of freedom, rather than the presence of an imaginary action alone, that gives rise to entanglement-induced decoherence.